Fig.1. (S.H. Naqib et al.) [Temperature dependence of electrical resistivity .....]

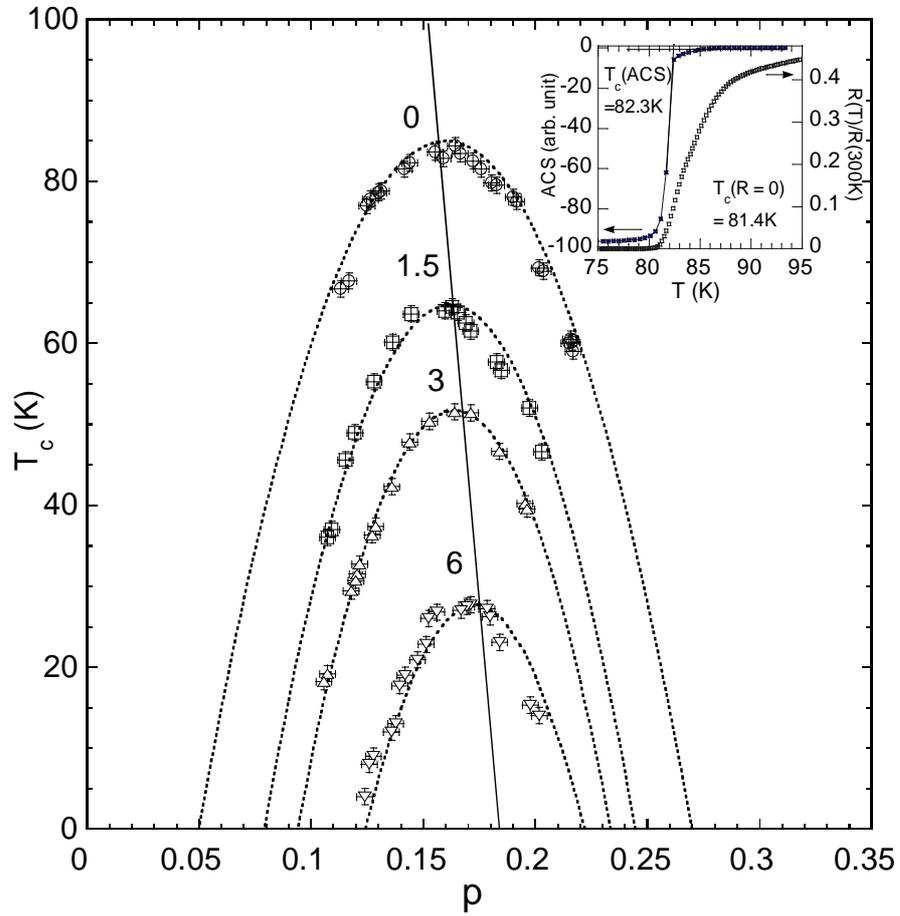

Fig.2. (S.H. Naqib et al.) [Temperature dependence of electrical resistivity....]

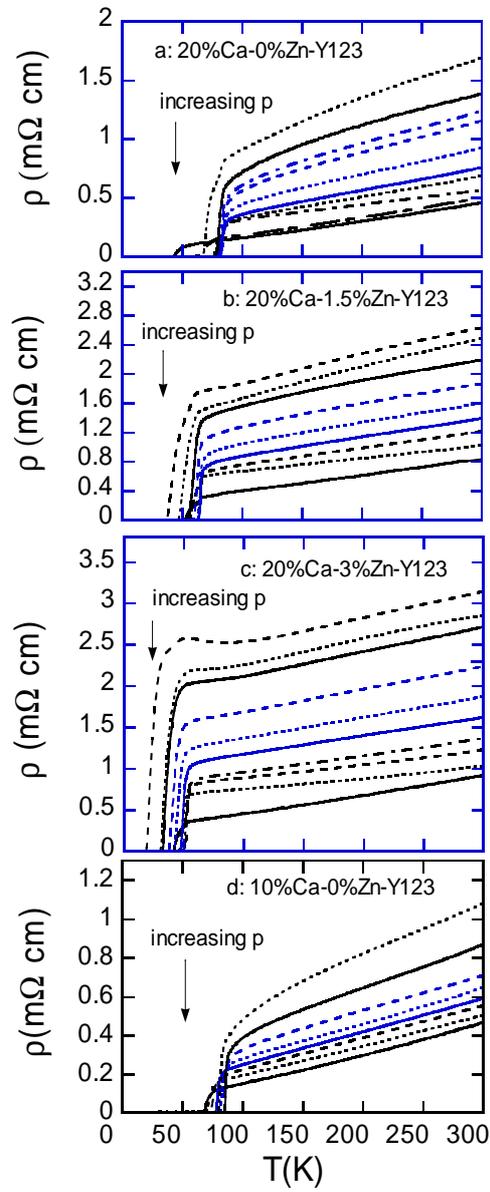

Fig.3. (S.H. Naqib et al.) [Temperature dependence of electrical resistiviy ....]

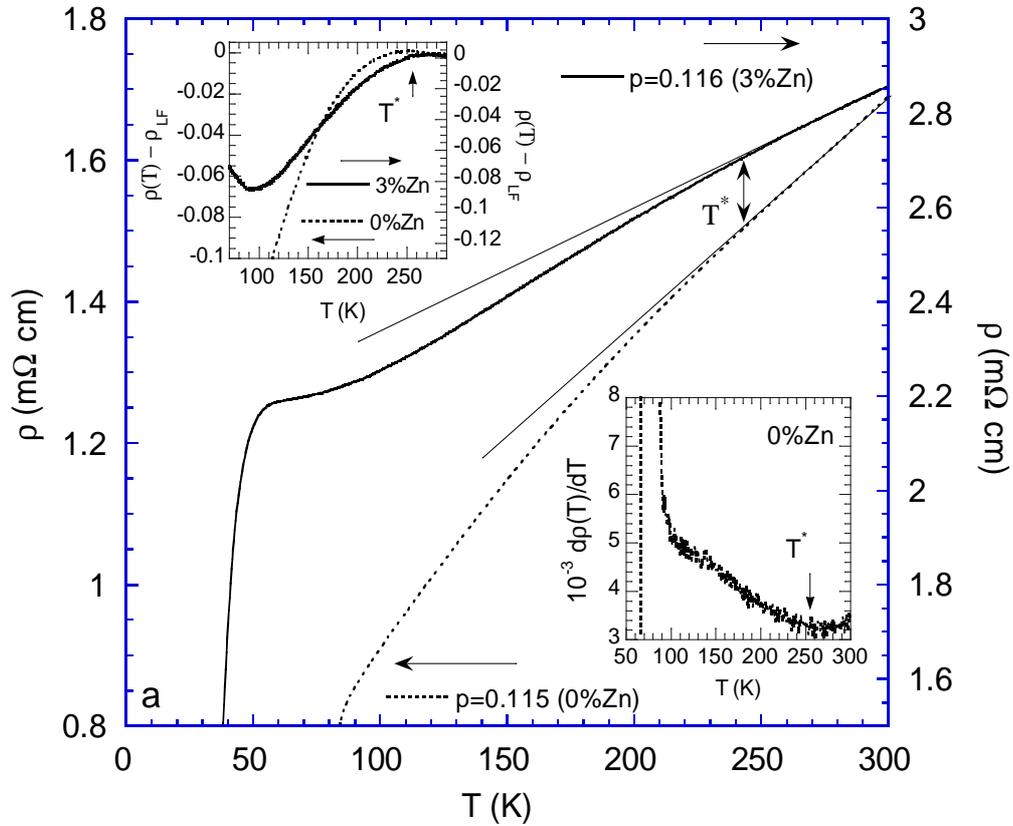

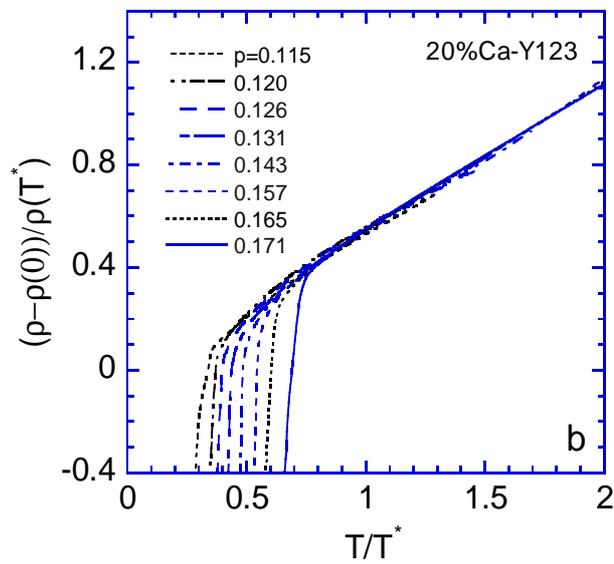

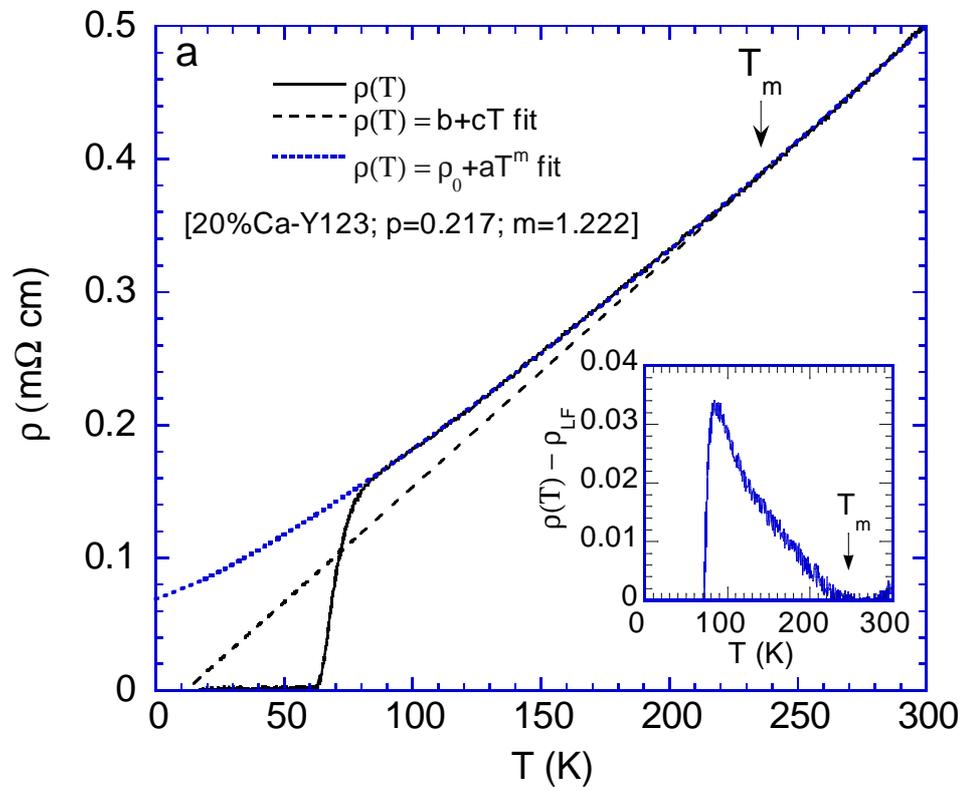

Fig.4b. (S.H. Naqib et al.) [Temperature dependence of electrical resistivity.....]

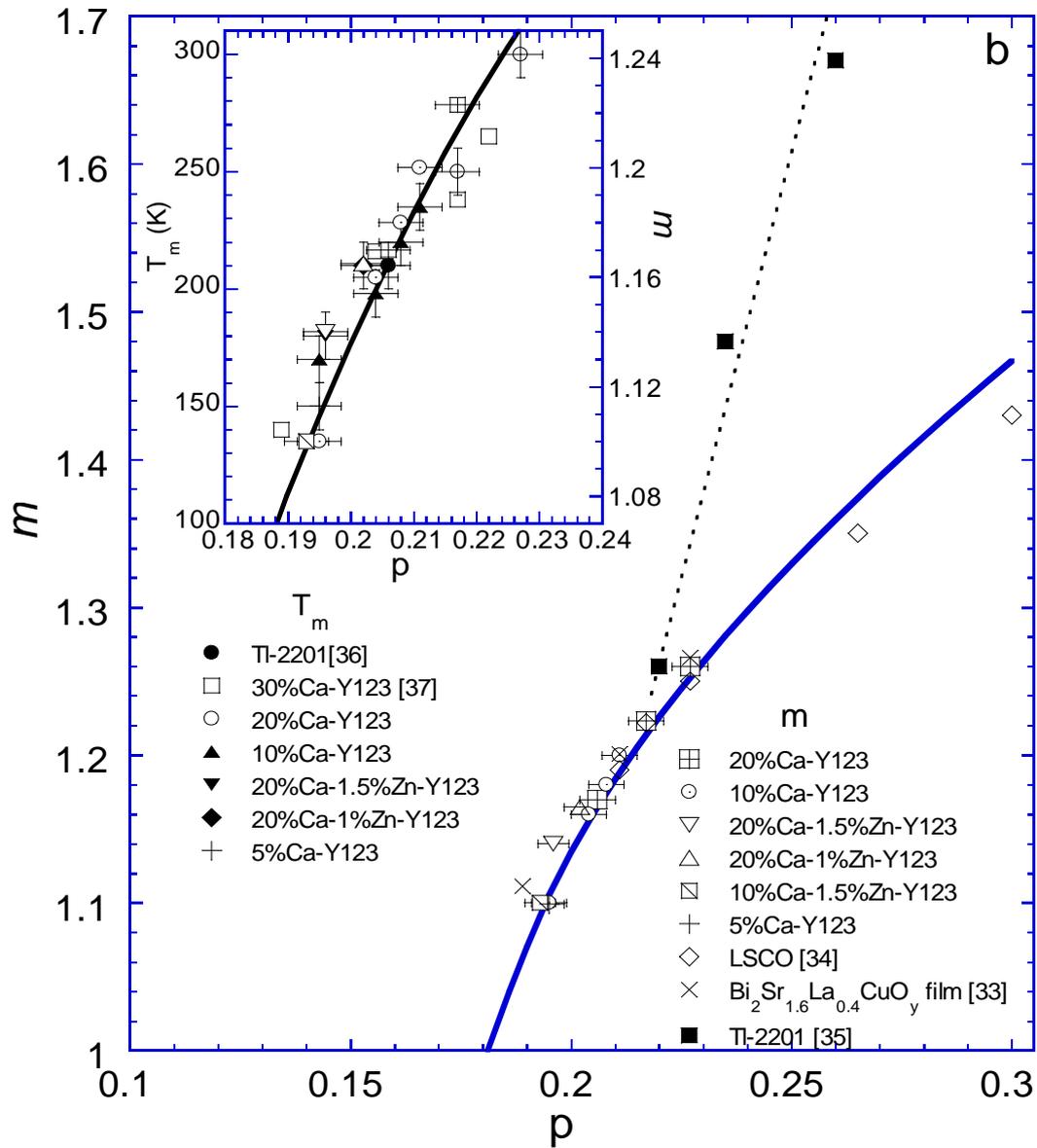

Fig.5. (S.H. Naqib et al.) [Temperature dependence of electrical resistivity ….]

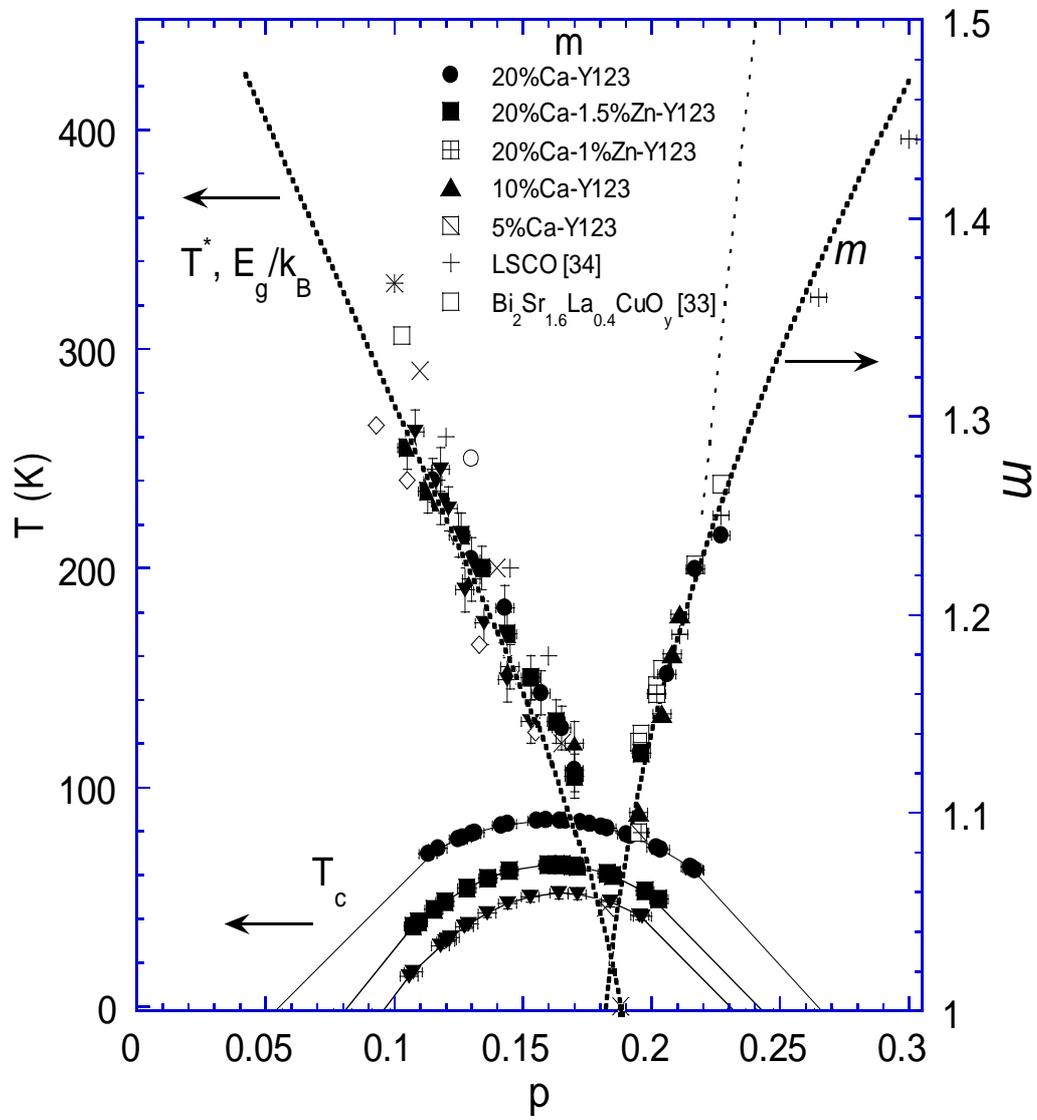

# Temperature dependence of electrical resistivity of high-$T_c$ cuprates - from pseudogap to overdoped regions

**S.H. Naqib[1*], J.R. Cooper[1], J.L. Tallon[2], and C. Panagopoulos[1]**

[1]IRC in Superconductivity and Dept. of Physics, Cambridge University, Madingley Road, Cambridge CB3 OHE, UK

[2]MacDiarmid Institute for Advanced Materials and Nanotechnology, Victoria University and Industrial Research Ltd.,

P.O. Box 31310, Lower Hutt, New Zealand

## Abstract

The effects of planar hole concentration, p, and in-plane disorder, Zn (y), on the DC resistivity, $\rho(T)$, of sintered samples of $Y_{1-x}Ca_xBa_2(Cu_{1-y}Zn_y)_3O_{7-\delta}$ were investigated over a wide doping range by changing both the oxygen deficiency ($\delta$) and Ca content (x). From the $\rho(T,p)$ data we extracted characteristic crossover temperatures on the underdoped and overdoped sides, $T^*$ and $T_m$ respectively, above which $\rho(T)$ is linear. We compare our results with a number of other polycrystalline, thin film and single crystal cuprate superconductors and find similar behavior in the p-dependence of $T^*(p)$, $T_m(p)$, and the resistivity exponent, $m(p)$, in fits to $\rho(T) = \rho_0 + aT^m$ on the overdoped side. Our findings points towards the possible existence of a quantum critical point (QCP) at the doping p=0.19 ± 0.01.

*PACS numbers*: 74.25.Dw, 74.25. 74.62.Dh, 74.72.-h

*Keywords*: Pseudogap; Quantum critical point

## 1. Introduction

The normal-state (NS) and superconducting (SC) properties of high-$T_c$ cuprates are extremely sensitive to stoichiometry and the number (p) of added hole carriers per copper oxide plane. In recent years one of the most widely studied phenomena in the physics of high-temperature superconductors (HTS) is the so-called pseudogap (PG) state [1,2] which is observed over a doping range extending from the underdoped (UD) to the slightly

*Corresponding author: Tel.: +44-1223-337049; Fax: +44-1223

-337074; E-mail: shn21@hermes.cam.ac.uk

overdoped (OD) region. In the PG-state various NS and SC anomalies are observed which can be interpreted in terms of a reduction in the effective single particle density of states [3]. Existing theories of the PG, which is believed to be an essential feature of HTS physics, can be classified broadly into two categories. The first is based upon a model of Cooper pair formation well above the SC transition temperature, $T_c$, with long-range phase coherence appearing only at $T \leq T_c$ [4-6]. The second assumes that the appearance of the PG is due to fluctuations of some other type, which compete or coexist with SC. These include antiferromagnetic fluctuations, charge density waves, structural phase transition or electronic phase separation on a microscopic sale (e.g., the stripe scenario) [7-10]. Within this second category the concept of a quantum critical point (QCP) has been mooted to explain the

HTS phase diagram [11-13] though its confirmation remains inconclusive.

The evolution of the resistivity, $\rho(T,p)$, provides a way of establishing the T-p phase diagram and can give valuable information about the different crossover regions in the vicinity of a possible QCP [11-13]. In this paper we report a systematic study of the transport properties of the SC compound $Y_{1-x}Ca_xBa_2(Cu_{1-y}Zn_y)_3O_{7-\delta}$. We have measured $\rho(T,p)$, the room-temperature thermopower, S[290K], and the AC susceptibility (ACS) of a series of sintered samples with different levels of Zn, Ca, and oxygen contents. Pure $YBa_2Cu_3O_{7-\delta}$ (Y123) with full oxygenation ($\delta=0$) is slightly OD and further overdoping can only be achieved by substituting $Y^{3+}$ by $Ca^{2+}$, while Zn substitution at the in-plane Cu(2) sites increases the planar impurity scattering. Here the doping level remains nearly independent of Zn content enabling one to separate the effects of doping and disorder on various NS and SC properties [14,15]. From the analysis of $\rho(T,p)$ we extract UD and OD crossover temperatures, $T^*$ and $T_m$ respectively, above which $\rho(T)$ exhibits linear T-dependence and we also determine the p-dependent exponent 'm' in fits to $\rho(T) = \rho_0 + aT^m$ on the overdoped side.

The main observations of the present study are that (i) the characteristic quantities $T^*(p)$, $T_m(p)$ and $m(p)$ remain independent of the disorder; and (ii) both $T^*(p)$ and $T_m(p)$ extrapolate to 0 K at $p = 0.19 \pm 0.01$, on the UD and the OD sides respectively, consistent with a possible QCP at $p_{crit} = 0.19 \pm 0.01$, where the exponent $m \rightarrow 1$ [12] over the whole temperature range. We have compared our values of $T^*(p)$, $T_m(p)$, and $m(p)$ with those of a number of other HTS and found similar qualitative behavior, irrespective of widely different chemical compositions and physical states (polycrystalline, c-axis oriented thin films, and bulk single crystals).

## 2. Experimental details and results

Polycrystalline single-phase samples of $Y_{1-x}Ca_xBa_2(Cu_{1-y}Zn_y)_3O_{7-\delta}$ were synthesized by standard solid-state reaction methods using high-purity ($> 99.99\%$) powders. The details of sample preparation and characterization can be found elsewhere [16,17]. Because of the strong p-dependence of various NS and SC properties (especially near $p_{crit}$) [11,18-20] it is important to determine p as accurately as possible. We have estimated p from the room temperature thermopower, S[290K] using the correlation of Obertelli et al. [21]. A previous study showed that S[290K] does not vary much with Zn content in Zn-substituted Y123 for fixed values of $\delta < 0.5$ [22]. We find the same trend in Zn-Ca-Y123. Accordingly, in this doping range S[290K] is still a good measure of p even in the presence of strong in-plane scattering by Zn. For Zn-free samples we have also calculated p using the $T_c(p)$ relation [23], given by

$$T_c(p)/T_{cmax}(p_{opt}) = 1 - Z(p - p_{opt})^2 \qquad (1)$$

with the usual values $Z = 82.6$ and $p_{opt} = 0.16$ (for maximum $T_c$) [23]. Using p from S[290K] and the experimental values of $T_c$ we obtained a very good fit of $T_c(p)$ with eq. (1) for the Zn substituted samples (see Fig.1). The values of $T_{cmax}$, Z, and $p_{opt}$, respectively, are 0% Zn: 84.5 K, 82.6, and 0.16; 1.5% Zn: 65.5 K, 146, and 0.162; 3% Zn: 51.8 K, 205, and 0.165; and 6% Zn: 27.5 K, 410, and 0.173. This systematic shift of $p_{opt}$ towards higher values with increasing Zn (y) is of particular importance. Fig.1 shows that superconductivity is at its strongest at p ~ 0.185 (i.e., the largest amount of Zn is needed to suppress superconductivity at this hole concentration), and as noted previously [11,20], this remains the last point of superconductivity at a critical Zn concentration. Notably, this doping level p ~ 0.19 is where the PG vanishes quite abruptly [11,18].

The hole content was varied either by changing the amount of Ca (x) or the oxygen deficiency ($\delta$) by quenching the samples into liquid nitrogen from different temperatures and oxygen partial pressures. Values of $\delta$ determined from the weight changes agree very well with an earlier study [18]. We have obtained $T_c$ from both resistivity and low field ($H_{rms} = 0.1$ Oe; f = 333.3 Hz) ACS measurements. $T_c$ was taken at zero resistance (within the noise level of $\pm 10^{-6}$ $\Omega$) and at the point where the line drawn on the steepest part of diamagnetic ACS curve meets the T-independent base line associated with negligibly small NS signal (see the inset of Fig.1). $T_c$ obtained from these two methods agrees within 1K for most of the samples.

High density (5.7 to 6.0 gm/cm$^3$) sintered bars were used for four-terminal

ρ(T) measurements with an ac current of 1 mA (77 Hz). Typical contact (silver paste) resistances were below 2Ω and transformers were used to minimise electrical noise. Fig.2 shows ρ(T) data for some of the samples used in the present study. We have used two methods to determine the PG temperature, $T^*$. As seen from Fig.3, representative plots of dρ(T)/dT vs. T and ρ(T) - $ρ_{LF}$ vs. T yield very similar $T^*$ values (within 5 K) ($ρ_{LF}$ is a linear fit $ρ_{LF}$ = b+cT, in the high-T linear region of the ρ(T), above T > $T^*$). Using $T^*$/T as a scaling parameter, it is possible to normalize ρ(T). The result of the scaling is shown in Fig.3b, where, leaving aside the SC transitions, all resistivity curves collapse reasonably well on to one p-independent universal curve over a wide temperature range. Similar scaling properties have been reported previously for the Hall effect [24], resistivity [25,26], magnetic susceptibility [27], specific heat [27], thermopower [28] and Knight shift [29] measurements. It is important to notice that, similar to other studies [30], Zn does not change the $T^*$(p) values but it suppresses $T_c$(p) very effectively (see Figs.3a, 2a, and 2c), this has often been stated as evidence for the non-SC origin of the PG [31].

A crossover temperature, $T_m$, can also be defined for the OD samples. Here a common feature is the increase in the positive curvature of ρ(T) with progressive doping and decreasing temperature. We define $T_m$ as the temperature below which ρ(T) starts to rise above the high-temperature linear fit (see Fig. 4a). Although this procedure gives consistent results for many samples, we urge caution in using linear fits to estimate $T_m$ and $T^*$. Namely, model calculations show that the validity of this method depends on the upper temperature limit of the available ρ(T) data [31] and therefore data up to T ~ 400 K are preferred when possible. For our sintered samples, reliable ρ(T)-data could not be obtained for T > 325 K due to the degradation of the grain boundaries probably because of the loss of minute quantities of oxygen there. Similar effects have been reported in previous studies [32]. For OD samples we have also determined the p-dependence of the exponent, $m$, in fits to $\rho(T) = \rho_0 + aT^m$. A single exponent fit describes the $\rho(T)$ of our samples over a wide temperature range (see Fig.4a). Since both $T_m$ and $m$ are measures of the deviation of $\rho(T)$ from linearity, we plot $T_m$ and $m$ vs. p in Fig.4b (see the inset). It is not surprising to find that $T_m$ and $m$ have similar p-dependence (thus indirectly confirming our values of $T_m$). There have been earlier studies [33-35] of the p-dependence of the resistivity exponent, $m$, for other HTS, and in Fig.4b we have included those for comparison. We also show the extracted $T_m(p)$ for $Tl_2Ba_2CuO_{6+\delta}$ (Tl-2201) single crystals and 30%Ca-Y123 c-axis oriented thin films from refs.[36] and [37] respectively. For these compounds, p was obtained using eq.(1) from the reported values of $T_c$. With the possible exception of Tl-2201 for p > 0.22, $m(p)$ seems to show universal behavior irrespective of the crystalline states and chemical composition (however the lack of $\rho(T)$ data for HTS cuprates with p ≥ 0.23 for compounds other than Tl-2201 and $La_{2-x}Sr_xCuO_4$ demands caution when talking about *exception*). Finally, in Fig.5 we construct an HTS "phase diagram". Fig.5 also shows $T^*(p)$ and the PG energy scale, $E_g$, from a number of different sources [11,25,33,34,38] using different experimental techniques. The values of $E_g(p)/k_B$ are from specific heat measurements [11] and an $^{89}Y$ NMR study [38], and $E_g(p)/k_B \sim \theta T^*(p)$, where $\theta$ (>1) is a proportionality constant [31]. In all cases $T^*(p)$ and $T_m(p)$ (and hence $m(p)$) exhibit similar behavior and both temperature scales → 0K at p = 0.19 ± 0.01, resembling behaviour near a QCP [11,12].

## 3. Discussion

The search for an underlying quantum phase transition in cuprate HTS and in heavy fermion compounds [39] has been motivated by the observation of various anomalous NS properties which deviate from a conventional Fermi liquid picture over much of the phase diagram. Due to the presence of competing order parameters in the vicinity of a QCP, the nature of the low-energy excitations are nontrivial and could lead directly to an explanation of the non-Fermi liquid behavior of the normal state metallic phase [11-13], scaling behaviour of the spin susceptibility and transport properties [24-29] and the overall generic behavior of cuprate HTS [11-13].

The effect of disorder (Zn and Ca) on $T^*(p)$ and $T_m(p)$ is insignificant (at least for the concentration levels shown in Fig.5). We have used $\rho(T)$ data for moderate Zn contents ($y \leq 1.5\%$) to obtain $m$ or $T_m$, since Zn tends to localize low-energy quasiparticles and induce a rise in the resistivity at low temperatures. This effect increases with decreasing p, and we only show Zn-substituted data for p > 0.19 in the OD side, where localization effects do not hamper determination of $T_m$ and $m$. With increasing p, $m(p)$ increases, even though, for $La_{2-x}Sr_xCuO_4$ (LSCO), at p ~ 0.30, $m$ ~ 1.45, and is still substantially lower than 2 (the canonical low-T Fermi liquid value with dominant electron-electron scattering). At this point we would like to mention that the single particle density of states, $N(\varepsilon)$, seems to pile up near the Fermi energy in OD cuprates [17,27,31,40] because the electronic specific heat coefficient, $\gamma$, can generally be fitted to the form $\gamma_0 + \gamma_1/(T+\lambda)$ (where $\gamma_0$, $\gamma_1$, and $\lambda$ are positive constants). $\gamma$ is related to the average $< N(\varepsilon) >_T$ over an energy region $\varepsilon \pm 2k_BT$. Such a T-dependence could significantly reduce the resistivity exponent. The precise effect is model dependent (e.g., it depends on how the Fermi velocity changes and how the final density of states affects the scattering rate), but generally speaking, at any given temperature, larger values of $<N(\varepsilon)>_T$ will increase the resistivity. Assuming $\rho(T) \propto <N(\varepsilon)>_T T^2$, the experimental $<N(\varepsilon)>_T$ for LSCO (p = 0.30) [40], reduces m from 2 to 1.7 for a model Fermi liquid system with dominant particle-particle scattering. The spin susceptibility, $\chi(T,p)$ (another measure of $<N(\varepsilon)>_T$), of Tl-2201 [35] shows a much weaker T-dependence than that of LSCO [27] in the OD side at the same p value, and the enhancement of the spectral weight near the Fermi level is much lower (see refs. 27 and 35). For example, for Tl-2201 at p ~ 0.25, $\chi(20K)/\chi(400K)$ ~ 1.146 and $\chi(20K)$ - $\chi(400K)$ ~ 0.30 ($10^{-4}$ emu/mol-Cu) [35], whereas for LSCO at p ~ 0.24, $\chi(20K)/\chi(400K)$ ~ 1.44 and $\chi(20K)$ - $\chi(400K)$ ~ 0.60 ($10^{-4}$ emu/mol-Cu) [27]. This could be a possible reason for substantially larger $m$(p) for Tl-2201 compared with LSCO in the heavily OD region.

Recently Hori et al. [41] studied the pressure dependence of $\rho_{ab}(T)$ of LSCO single crystals. At p = 0.22, superconductivity was almost completely suppressed at a pressure of 8 GPa. Surprisingly, even in this case where p is unchanged, $m$ (~ 1.2, at ambient pressure) increases monotonically with increasing pressure as superconductivity is weakened and reaches ~ 1.4 at a pressure of 8 GPa, nearly the same value for the ambient pressure $m$ of the non-SC LSCO sample with p = 0.26 [41]. This apparently indicates that a more Fermi liquid-like ground state formed at the expense of superconductivity, consistent with the findings of the present study. In a recent μSR study in the SC state of LSCO and $Ba_{2.1}Sr_{1.9}Ca_{1-x}Y_xCu_2O_{8+\delta}$, Panagopoulos et al. [42] found that at a critical doping, p ~ 0.19, there was an abrupt change in the magnetic spectrum and both the spin fluctuation temperature, $T_f$, and the spin glass temperature, $T_g$ (where fluctuations freeze), tend to zero at this doping. These results provide independent evidence for a quantum phase transition that separates the phase diagram of HTS into two distinct ground states [42].

The T-p phase diagram (Fig.5) bears a strong resemblance to the schematic T-x (x = p) phase diagram, derived by Varma [12], in the vicinity of a QCP. The PG-state corresponds to an "ordered" state [11-13] of some kind, though its nature has not been properly identified and remains controversial. Also, no thermodynamic evidence for any phase transition associated with $T^*$ has been observed [11]. However, very recently Kaminski et al. [43] reported the spontaneous breaking of time-reversal symmetry in the PG-state of $Bi_2Sr_2CaCu_2O_{8+\delta}$ in the ARPES spectrum. This symmetry breaking was only seen in the PG state in UD compounds and was absent in the OD samples [43].

The robustness and the universality of $T^*(p)$ and $m(p)$ or $T_m(p)$, with the notable possible exception of Tl-2201 in highly OD region, are striking. At the same time, $T^*(p)$ does not correlate with $T_c(p)$ in any way. For example, $T^*(p)$ of LSCO ($T_{cmax}$ = 38 K) is almost twice of that for Zn-Ca-Y123 ($T_{cmax}$ = 93 K), $T^*(p)$ of $Bi_2Sr_{1.6}La_{0.4}CuO_y$ (a single layer compound like LSCO) is comparable to that of Zn-Ca-Y123 but its $T_{cmax}$ = 30 K is comparable to LSCO. The only qualitative connection between $T_c(p)$ and $T^*(p)$ for all these compounds is that one grows at the expense of the other. These place a severe constraint on pictures where the PG is derived from SC fluctuations.

## 4. Conclusion

In summary, we have tried to elucidate the HTS phase diagram based on our $\rho(T)$ measurements on $Y_{1-x}Ca_xBa_2(Cu_{1-}$

$_y$Zn$_y$)$_3$O$_{7-\delta}$ over a wide range of doping and composition. We have identified crossover temperatures, $T^*(p)$ and $T_m(p)$ above which $\rho(T)$ becomes linear, and compared these values for other families of cuprates. The doping dependence of these characteristic parameters appears generic to many HTS and indicates strongly that $p = 0.19 \pm 0.01$, where the PG vanishes, is a fixed point, reminiscent of a QCP where $\rho(T)$ remains linear in the NS as $T \rightarrow 0K$.

## Acknowledgements

We thank J.W. Loram for helpful comments. S.H.N acknowledges financial support from the Commonwealth Scholarship Commission (UK), Darwin College, Cambridge Philosophical Society, Lundgren Fund, and the Department of Physics Cambridge University. C.P. acknowledges the Royal Society for financial support and JLT acknowledges support from the Marsden Fund.

Fig.1. $T_c$ versus p for 20%Ca-Y123, with different Zn concentrations (y) (shown in %). p was calculated from S[290K]. $T_c$(p)-fit (dotted lines) was drawn using p from S[290K]. Inset shows $T_c$ from R(T) and ACS measurements for a Zn-free sample (p = 0.171).

Fig.2. Some representative $\rho$(T) data (p values from top are: (a) 0.115, 0.131, 0.143, 0.157, 0.165, 0.171, 0.183, 0.193, 0.217, 0.227; (b) 0.108, 0.115, 0.127, 0.137, 0.147, 0.161, 0.167, 0.19, 0.201; (c) 0.107, 0.116, 0.121, 0.127, 0.135, 0.144, 0.152, 0.167, 0.185, 0.198; (d) 0.131, 0.147, 0.17, 0.183, 0.196, 0.204, 0.208, 0.213.

Fig.3. (a) Methods used for determining $T^*$ of $Y_{0.80}Ca_{0.20}Ba_2(Cu_{1-y}Zn_y)_3O_{7-\delta}$ (b) Normalized resistivity with $T^*$ as a scaling parameter for $Y_{0.80}Ca_{0.20}Ba_2Cu_3O_{7-\delta}$ ($\rho$(0) = b in $\rho_{LF}$, see text).

Fig.4. (a) Methods used for extracting $T_m$ and $m$. (b) Main part, $m$ vs. p (Inset, $T_m$ and $m$ vs. p). The solid line is drawn as a guide to the eye. The dotted line shows the $m$(p)-trend for Tl-2201 for p > 0.22 [35].

Fig.5. Phase diagram of HTS cuprates ($T_c$, $T^*$, and $E_g/k_B$: (●) 20%Ca-Y123, (■) 20%Ca-1.5%Zn-Y123, (▼) 20%Ca-3%Zn-Y123, (▲) 10%Ca-Y123,. (O), 20%Ca-Y123 (NMR) [38], (×) 20%Ca-Y123 (Specific heat) [11], (+) 0.5$T^*$ of LSCO [34], (□) $Bi_2Sr_{1.6}La_{0.4}CuO_y$ c-axis thin film [33], (◇) Y123 c-axis thin film [25]). The legends for $m$ are shown in the figure. The dotted line shows the $m$(p)-trend for Tl-2201 for p > 0.22 [35]. The dashed lines are drawn as guides to the eye. $T^*$(p) are from $\rho$(T) measurements.